\documentclass[onecolumn,
		nofootinbib
		]{revtex4}
\usepackage{graphicx,color}

\usepackage{amssymb}
\usepackage{amsbsy}
\usepackage{amsthm,mathrsfs,amsopn}
\usepackage{amsmath}

\theoremstyle{plain} 

\newtheorem{theorem}{Theorem}[section]

\newtheorem{remark}[theorem]{Remark}

\begin{document}

\title{Urban skylines from Schelling model}

\author{F. Gargiulo$^{1}$, Y. Gandica$^{1}$, T. Carletti$^{1}$}
\affiliation{
1. Department of Mathematics and Namur Center for Complex Systems - naXys, University of Namur, rempart de la Vierge 8, B 5000 Namur, Belgium}



\begin{abstract}
We propose a metapopulation version of the Schelling model where two kinds of agents relocate themselves, with unconstrained destination, if their local fitness is lower than a tolerance threshold. We show that, for small values of the latter, the population redistributes highly heterogeneously among the available places. The system thus stabilizes on these heterogeneous skylines after a long quasi-stationary transient period, during which the population remains in a well mixed phase.\\
Varying the tolerance passing from large to small values, we identify three possible global regimes: microscopic clusters with local coexistence of both kinds of agents, macroscopic clusters with local coexistence (soft segregation), macroscopic clusters with local segregation but homogeneous densities (hard segregation). The model is studied numerically and complemented with an analytical study in the limit of extremely large node capacity.
\end{abstract}
\maketitle

\section{Introduction}
Modern societies are often faced to segregation; dictated by race, religion, social status or incomes differences, it represents a major issue whose outcome can range from social unrest, to riots and possibly to civil wars. The understanding of the rise of such phenomenon has thus attracted a lot of attention from economists, politicians and sociologists~\cite{CutlerGlaeser1997,FagioloEtAl2007,FreyFarley1996,ConradtEtAl2009}. 

In a couple of papers written in the late 60s, Thomas Schelling~\cite{Schelling1969,Schelling1971} proposed a stylized model to describe the unset of segregation and pointed out a counterintuitive but widely observed result: a well integrated society can evolve into a rather segregated one even if at individual level nobody strictly prefers this final outcome. Even when individuals are quite tolerant to neighbors of their opposite kind, allowing them to relocate themselves to satisfy their preferences - namely maximize their perceived local fitness/utility - will make segregation to emerge as a global aggregated phenomenon not directly foreseen from the individual choices.

Since the pioneering works of Schelling the model has attracted the attention of the community of physicists and mathematicians, interested in its simplicity and in the emergent behaviors recalling models such as Ising and Potts ones~\cite{VinkovicKirman2006, DallAstaEtAl2008,Schulze2005,StaufferSolomon2007,GrauwinEtAl2009,RogersMcKane2012,RogersMcKane2011}. In its simplest form the model proposed by Schelling considers a population composed by two kinds of agents sitting on the top of a regular lattice (or a 1 dimensional ring), each site being able to receive at most one agent. The latter being allowed to hop to a new empty lattice site once unhappy, that is once the fraction of agents of her opposite kind in her Moore neighborhood, is larger than a given {\em tolerance threshold}. The striking result by Schelling is that segregation will emerge for tolerances slightly larger than $1/3$, well below the \lq\lq more natural bound\rq\rq $1/2$.

Several variations have been performed onto the original Schelling model (see~\cite{PancsVriend2007} for a review), for instance different ways of computing the agent utility~\cite{RogersMcKane2011} and rules according which agents do move (swap agents positions, restrict moves to increasing fitness locations~\cite{VinkovicKirman2006,DallAstaEtAl2008}, to fixed distance places or following the path of an underlying network~\cite{FagioloEtAl2007}). All such models contains essentially two parameters, the tolerance threshold and the fraction of empty spaces; the latter having received much more attention because it is responsible for a phase transition in the model~\cite{VinkovicKirman2006,DallAstaEtAl2008,GrauwinEtAl2009,RogersMcKane2011,RogersMcKane2012}.

Recently the Schelling model has been improved in realism by considering a metapopulation scheme~\cite{GrauwinEtAl2009,RogersMcKane2012,DurettYuan2014}. Nodes have a large but finite carrying capacity representing thus a bunch of residences in a district whose spatial extension can be neglected (well mixed population); happiness and thus willingness to move is thus conditioned by the fraction of agents of the opposite kind inside a given node, with respect to the nodes occupancy ($1$-node fitness)~\cite{DurettYuan2014}. 

\section{The Model}
In this paper we elaborate further in this direction by considering a metapopulation version of the Schelling model where two kinds of agents, say Red and Blue, can move across $N$ nodes, each of which can receive at most $L$ agents~\footnote{To make things simple we assumed a constant carrying capacity for each node, but of course one can improve the mode by considering a different value of $L_i$ for each node $i$.} (carrying capacity) at any given time. We assume there are in the model $\rho LN$ vacancies, i.e. empty spaces, and an equal number of Red and Blue agents.

The considered model resembles to the one proposed in~\cite{DurettYuan2014}, the main difference being that in our case agents don't have any information about the selected destination, therefore also moves that decrease or leave invariant the fitness are allowed, we define such case \emph{weak liquid} version of the Schelling model, the liquid case referring in the literature to agents' moves for which the fitness does not decrease~\cite{VinkovicKirman2006}. A second difference is that in~\cite{DurettYuan2014} the agent fitness is computed using single node informations, namely the number of agents in a given sites, on the contrary we reintroduce as Schelling originally did, the concept of spatial proximity~\cite{Schelling1971}, agent fitness takes into account the number of agents in a given node and in the neighboring ones, namely nodes at distance $1$ from the current node (local fitness).

Nodes are assumed to be arranged in regular lattices, as initially assumed by Schelling, but the model can be easily extended to complex networks. The local update rules is defined as follows. The neighborhood of an agent is given by the topological neighborhood of the node where she lives, including the latter. At each time step an agent is selected and her fitness is computed as the fraction of agents of the opposite kind of her, living in her neighborhood with respect to the total
number of agents living in the same neighborhood. Mathematically, assuming she is a Blue agent living in node $i$, then her fitness is given by: 
\begin{equation}
\label{eq:fitness}
{f}_i^B=\frac{\sum_{j\in i}n_j^A}{\sum_{j\in i}(n_j^B+n_j^A)}\, ,
\end{equation}
where $n_j^X$, $X=A,B$, is the number of agent of $X$-kind in node $j$, and we used the notation $j\in i$ to denote all nodes $j$ belonging to a neighborhood of node $i$, including the latter, that is the set of nodes at distance smaller or equal to $1$ from $i$.

As in the original Schelling model, agents are unhappy if their fitness is larger than a given {\em tolerance threshold}, $\epsilon\in(0,1)$, hereby assumed to be the same for all agents of both kinds; to reduce their uneasiness agents move by choosing uniformly at random another node $k$ not completely full, i.e. $n_k^A+n_k^B<L$:
\begin{equation}
\label{eq:faction}
\text{if ${f}_i^B>\epsilon$} \Rightarrow \text{agent $A$ leaves node $i$.}
\end{equation}
The case for a Red agent is similar.

One time step is the random selection with reinsertion of $(1-\rho)NL$ agents. We define the {\em convergence time}, to be the time needed for the system to reach the equilibrium, namely once no agents will move anymore.

\section{Results}
\label{sec:res}

We hereby present the numerical analysis of the proposed model once the underlying network is a regular lattice with periodic boundary conditions and each node has $4$ neighboring nodes. The system is initialized with $\rho NL$ vacancies, $(1-\rho)NL/2$ Red agents and the same number of Blue ones, uniformly random distributed among the $N=400$ nodes. The carrying capacity has been fixed to $L=100$ and we check that the initial conditions satisfy the local constraint $n_i^A+n_i^B\leq L$ for all $i$. Throughout the paper the emptiness has been fixed to $\rho=0.9$.

\subsection{Single node properties}
The aim of this section is to present the local properties of the system, namely at the level of single nodes. The metric we used is the {\em average value}, over all the nodes, of the node {\em magnetization}:
\begin{equation}
\label{eq:magnetization}
\langle\mu\rangle=\frac{1}{N}\sum_i\frac{|n_i^B-n_i^A|}{n_i^B+n_i^A}\, ,
\end{equation}
small values of $\langle\mu\rangle$ mean that, on average, each node is populated by the same number of agents of both kinds, while large values are associated to nodes filled with agents of only one kind.

\begin{figure*}
\centering
\includegraphics[width=1\textwidth]{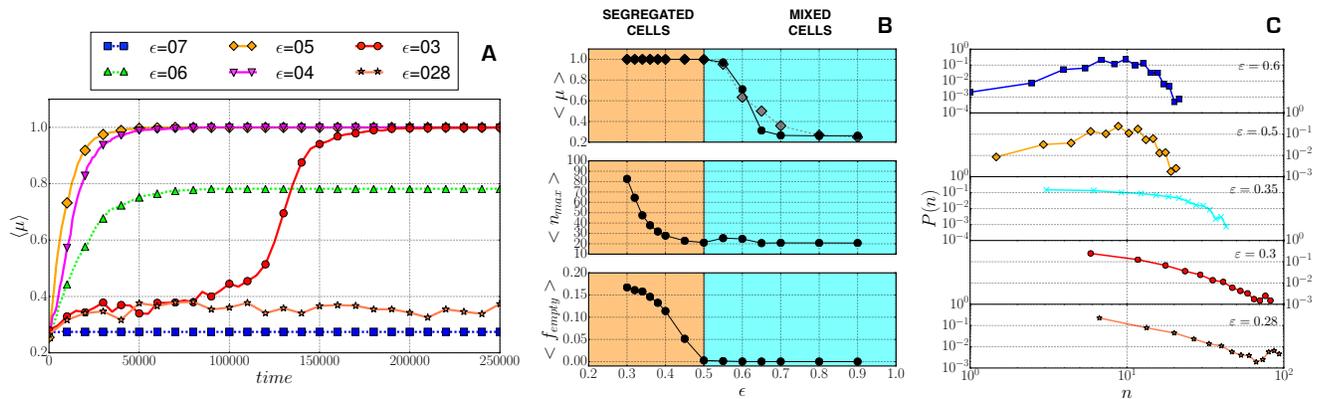}
\caption{\label{local}Single node behavior. Panel A: average magnetization $\langle\mu\rangle=\frac{1}{N}\sum_i\frac{|n_A^i-n_B^i|}{n_A^i+n_B^i}$ as a function of time for a single generic simulations. For $\epsilon\leq 0.5$ the magnetization goes to $1$ indicating the total segregation inside each node. Panel B: asymptotic average magnetization (upper plot), average maximal node population $\max_i(n_i^A+n_i^B)$ (middle plot) and fraction of empty nodes $f_{empty}$ (lower plot) as a function of $\epsilon$. For $\epsilon \leq 0.5$ the asymptotic average magnetization is constantly equal to $1$, showing the robustness of the local hard segregation against replicas. As $\epsilon\rightarrow 0$ agents of the same kind tend to select and fill all the compatible - with respect to the happiness - nodes, creating thus an important fraction of empty nodes. The black points correspond to the model with local fitness (namely distance-$1$ nodes) while the grey ones to the simplified model with $1$--node fitness (namely distance $0$ nodes). As we can see the magnetization behavior doesn't depend on the way the fitness is computed. Panel C: distribution of nodes total population ($n=n_A+n_B$) for few values of $\epsilon=0.28,0.3,0.35,0.5,0.7$.  Averages shown in Panels B and C have been obtained performing $100$ replicas of the model.}
\end{figure*}

For $\epsilon\leq 0.5$ we observe (see Fig.~\ref{local} panels A and B upper plot) that asymptotically $\langle\mu\rangle\rightarrow1$, meaning that the system stabilizes into a frozen state where {\em local segregation} is present in all nodes: each node contains only agents of one kind. Let us observe (see Fig.~\ref{local} B upper plot) that the same behavior is also present in the simplified model where the fitness is calculated on the single node, as done in~\cite{DurettYuan2014}, and thus it is intrinsic to the displacement dynamics and not to the way the agent fitness is computed.
We also notice (see Fig.~\ref{local} panels A and Fig.~6 that as $\epsilon$ decreases toward zero, the time needed to reach the frozen state gets longer, going to infinity in the limit  $\epsilon\rightarrow 0$ . The system exhibits thus a transient phase where it remains stuck for very long time into a {\em quasi-stationary non-segregated state} (see red circles and orange stars curves in Fig.~\ref{local}A).

A second fundamental self-organized phenomenon emerges for $\epsilon\leq 0.5$, the initial homogeneously distributed population organizes itself into an heterogeneous state across the network nodes (Fig.~\ref{local} panel C lower plots): most of the nodes contain $\sim 10$ agents, while very few nodes have as much as $\sim 100$ agents, recall that $L=100$ is the maximum node capacity. 

Observe that such asymptotic distribution is correlated with the time the system spends in the quasi-stationary non-segregated state, the longer this time the more the population distribution across nodes moves from a Poissonian distribution (Fig.\ref{local} panel C upper plots) to a power law (Fig.~\ref{local} panel C lower plots). The maximal node population $n_{max}=\max_i(n_i^A+n_i^B)$ increases for $\epsilon\rightarrow 0$ (Fig.~\ref{local} panel B middle plot). Notice that the maximal node capacity ($L=100$) is never reached. At the same time we observe the formation of a relevant fraction of completely empty nodes, i.e. nodes for which $n_i^A+n_i^B=0$ (Fig.~\ref{local} panel B lower plot). 

The transition to the local segregation at $\epsilon=0.5$ is an important issue of the metapopulation version of the Schelling model that can be easily explained once the fitness depends only on one node, with the following argument. A mixture of agents of both kinds in each node, determines a global happy configuration if and only if:
\begin{equation}
\forall i\quad \frac{n_A^i}{n_A^i+n_B^i}\leq\epsilon \qquad \frac{n_B^i}{n_A^i+n_B^i}\leq\epsilon\, ,
\end{equation}
but it is straightforward to observe that this equation admits a solution, for which $n_A^i>0$ and $n_B^i>0$, only for $\epsilon>1/2$. Hence for $\epsilon \leq1/2$ the above relations cannot be satisfied and thus unhappy agents start to move, increasing the happiness of agents of the same kind and decreasing the happiness of agents of opposite kind. The net result of such behavior is segregation. 

\subsection{Global properties}
\label{ssec:global}

To analyze the spatial structures, thus beyond the single node, we define a {\em cluster} based on node's majority, more precisely two linked nodes belong to the same cluster if they both are characterized by the majority of agents of the same kind~\footnote{Excluding the case where there is not strict majority in a node, this definition can be restated as: $i$ and $j$ belong to the same cluster if $(n^A_i-n^B_i)(n^A_j-n^B_j)>0$.}:
\begin{equation}
\label{eq:cluster}
\text{$i,j\in C$ if \{$n_i^A>n_i^B$ and $n_j^A>n_j^B$\} or \{$n_i^A<n_i^B$ and $n_j^A<n_j^B$\}}\, .
\end{equation}
If a node share the same (positive) number of Red and Blue agents, then it will be considered part of the {\em interface}. An edge between two neighboring nodes $i$ and $j$ is considered interface if $(n^A_i-n^B_i)(n^A_j-n^B_j)\leq 0$. Hence a cluster is made by nodes while an interface can contains both nodes and edges among them.

\begin{figure*}[bt]
\centering
\includegraphics[width=1\textwidth]{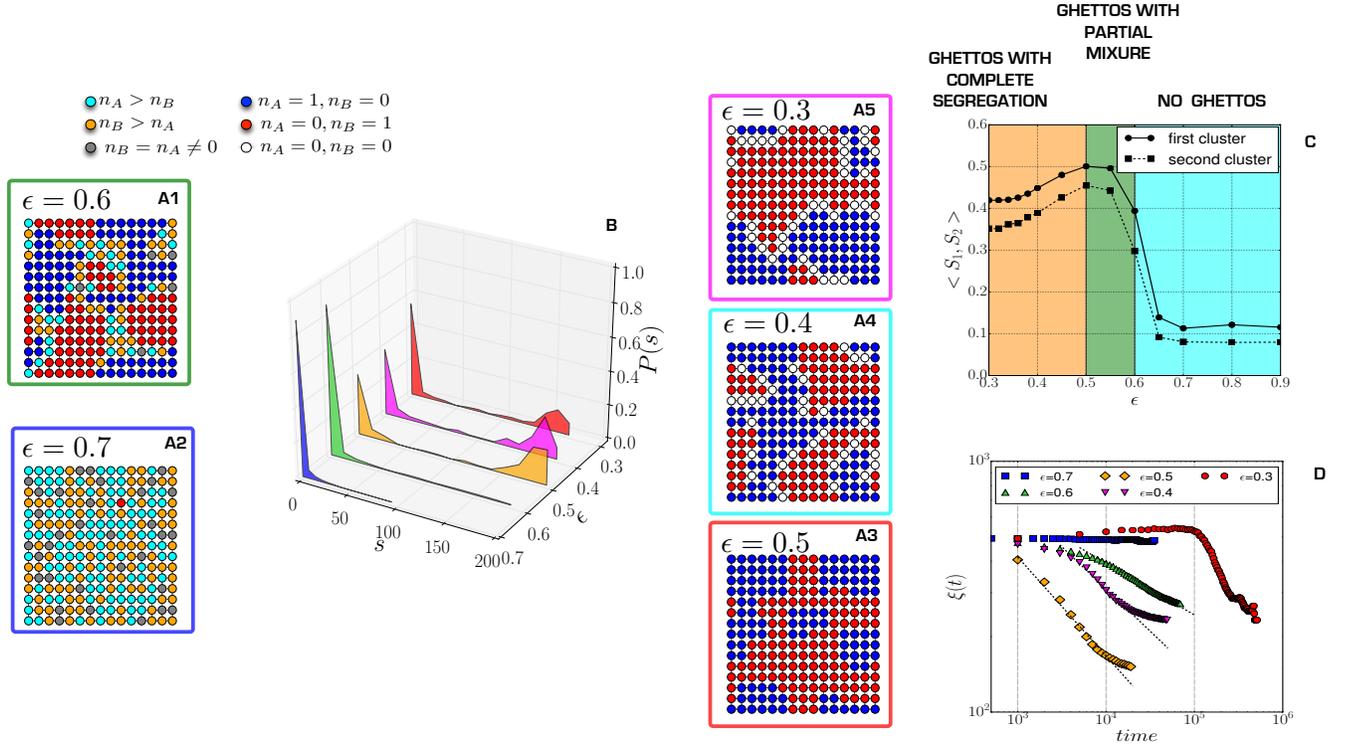}
\caption{\label{global}Global behavior. Panels A: Nodes occupancy at equilibrium, each node has been colored according to the majority of agents in it: white empty nodes, blue only B agents, R only red agents and shades for intermediate cases. Panel B: Cluster size distributions for several values of $\epsilon=0.3,0.4,0.5,0.6,0.7$. Panel C: First and second largest cluster as a function of $\epsilon$. Panel D: Interface size as a function of time  for several values of $\epsilon=0.3,0.4,0.5,0.6,0.7$. Results presented in panels B, C and D are based on $100$ replicas of the simulations.}
\end{figure*}

This indicator allows us to show that (Fig.~\ref{global} panels B and C) for $\epsilon\leq 0.6$ macroscopic percolating clusters are always formed, the average size of the largest cluster oscillating between $0.4N$ and $0.5N$ and at the same time the average sizes of the first and second largest cluster added together cover more than $75\%$ of the available nodes. This critical threshold is the same observed in the original Schelling model. Notice that for $\epsilon=0.5$ the first and second cluster cover almost completely the lattice, containing more than $95\%$ of nodes. From the behavior of $\langle S_{max}\rangle$ for $0.5<\epsilon\leq 0.6$ and the results of Fig.~\ref{local} Panel B, one can conclude that the largest clusters contain a majority of agents of the same kind, but different agents can coexist in the same node. We call this scenario \emph{soft segregation} because it allows a small mixing in the population. For $\epsilon\leq 0.5$ each spatial cluster contains only one kind of agent, resulting in a \emph{hard segregation} of the population. For $\epsilon\rightarrow 0$ clusters of empty cells are created increasing the distance between the two monochromatic structures by interposing interfaces (Fig.~\ref{local} panels B lower plot and white nodes in Fig.~\ref{global} panels A3, A4 and A5).

Looking at the temporal growth of the interfaces (Fig.~\ref{global} panels D), we can observe another remarkable self-organized phenomenon: for $0.3<\epsilon\leq0.6$ the size of the interface, monotonically decreases in time as $t^{-1/z}$ ($z=4$ for $\epsilon=0.6$ and $z=3$ for $\epsilon\geq 0.5$).  This is the typical signature of a coarsening phenomenon, that has already been observed in the classical Schelling model~\cite{DallAstaEtAl2008}. 
On the other side, for low values of the tolerance where the system remains for long time in the quasi-stationary state (for instance $\epsilon\leq 0.3$), the size of the interface has a slowly increasing phase (red curve Fig.~\ref{global} panel D), corresponding to the formation of temporal segregated domains followed by an abrupt decrease when the system reaches the local segregation equilibrium. In this case the domain formation mechanism cannot be ascribed to the coarsening framework: the spatial segregation indeed is reached through a mechanism of aggregation of agents around the high density instabilities that allow the system to exit the quasi-stationary state. Once the clusters are formed, the complete equilibrium is reached as the interface between Blue and Red zones becomes empty (See Fig.~7 for a detailed explanation). 

While the creation of heterogeneity distribution of agents among nodes is due to the metapopulation mechanism, the formation of the monochromatic clusters was already present in the classical Schelling model and thus is mainly due to the behavioral rules of the latter. We can therefore imagine the process at play in our model as the combined outcome of a local birth and death process (migration in-out) on the single node and a coarse-grained node dynamics on the lattice.

\section{An analytical simplified model}
To gain insight into the behavior of the previously presented model, we hereby introduce a model ables to capture the main behavior of the metapopulation Schelling model, but simple enough to be analytically tractable. Because we consider the case for extremely large $L$, our model complements the one proposed in~\cite{DurettYuan2014} devoted to the case $L=2$.

Using the notations introduced before the state of the system is thus completely characterized by the knowledge of $(\vec{n}^A(t),\vec{n}^B(t))$, being $\vec{n}^A(t)=(n^A_1(t),\dots,n^A_N(t))$ and $\vec{n}^B(t)=(n^B_1(t),\dots,n^B_N(t))$. For a sake of simplicity we decided to compute the fitness using only the information from a single node, Eq.~\eqref{eq:fitness} with $j=i$. The system evolution is done as previously and we still use the weak liquid version, an unhappy agent will move to an uniformly randomly chosen new node, provide there is enough space there. 

The model is thus intrinsically stochastic and hence it can be described by the probability $P(\vec{n}^A,\vec{n}^B,t)$ to be at time $t$ in state $(\vec{n}^A,\vec{n}^B)$, whose evolution is dictated by the master equation:
\begin{eqnarray}
\label{eq:ME}
P(\vec{n}^A,\vec{n}^B,t+1)&=&P(\vec{n}^A,\vec{n}^B,t)+\\&+&\hspace{-2em}\sum_{(\vec{n}^{A^\prime},\vec{n}^{B^\prime})} \hspace{-1em}\big[ T(\vec{n}^A,\vec{n}^B\vert\vec{n}^{A^\prime},\vec{n}^{B^\prime})P(\vec{n}^{A^\prime},\vec{n}^{B^\prime},t)\notag\\
&-&T(\vec{n}^{A^\prime},\vec{n}^{B^\prime}\vert\vec{n}^A,\vec{n}^B)P(\vec{n}^A,\vec{n}^B,t)\big]\notag\, ,
\end{eqnarray}
being $T(\vec{n}^{A^\prime},\vec{n}^{B^\prime}\vert\vec{n}^A,\vec{n}^B)$ the {\em Transition probability} to pass from state $(\vec{n}^A,\vec{n}^B)$ to the new compatible one $(\vec{n}^{A^\prime},\vec{n}^{B^\prime})$. For incompatible states we  set $T(\vec{n}^{A^\prime},\vec{n}^{B^\prime}\vert\vec{n}^A,\vec{n}^B)=0$. 

The non zero transition probabilities can be computed using the rules previously defined (see Appendix\ref{sec:analytical} for a detailed account of the transition probabilities calculation), for instance the transition probability that an $A$ agent moves from node $i$--th to node $j$--th is given by:
\begin{eqnarray*}
&&T_1(n^A_i-1,n^A_j+1\vert n^A_i,n^A_j)= \\&=&\frac{1}{N}\frac{n^A_i}{n^A_i+n^B_i}\Theta\left(\frac{n^B_i}{n^A_i+n^B_i}-\epsilon\right)\frac{L-n^A_j-n^B_j}{L}\, .
\end{eqnarray*}
The function $\Theta(x)$, defined to be $1$ if $x>0$ and $0$ otherwise, translates the willingness of an unhappy agent to move; smoother versions can be used as well. Let us observe that to lighten the notation we wrote only the variables whose values change because of the transition.

From the master equation one can compute the time evolution of some relevant quantities, for instance the average number of $X$--agents in every node $i$ at time $t$, $\langle n^X_i\rangle(t)=\sum_{\vec{n}^X}n^X_iP(\vec{n}^A,\vec{n}^B,t)$, $X=A,B$.

Using the expression for the transition probabilities Eq.~[B3] and assuming correlations can be neglected, i.e. $\langle (n^A_i)^2\rangle\sim \langle n^A_i\rangle^2$, we obtain a system of finite differences describing the evolution of $\langle n^A_i\rangle$ and $\langle n^B_i\rangle$.

To go one step further, we introduce the average fraction of $A$ and $B$ in each node, $\alpha_i={\langle n^A_i\rangle}/{L}$ and $\beta_i={\langle n^B_i\rangle}/{L}$, we rescale time by defining $s=t/L$ and finally we assume each node to have an infinite large carrying capacity (see Appendix\ref{sec:analytical} for a more detailed discussion):
\begin{eqnarray}
\label{eq:alphat1}
\frac{d\alpha_i(s)}{ds}&=&-{\rho}\frac{\alpha_i}{\alpha_i+\beta_i}\Theta\left(\frac{\beta_i}{\alpha_i+\beta_i}-\epsilon\right)\notag\\&+&\frac{1-\alpha_i-\beta_i}{N}\sum_{j}\frac{\alpha_j}{\alpha_j+\beta_j}\Theta\left(\frac{\beta_j}{\alpha_j+\beta_j}-{\epsilon}\right)
\end{eqnarray}
and
\begin{eqnarray}
\label{eq:betat1}
\frac{d\beta_i(s)}{ds}&=&-{\rho}\frac{\beta_i}{\alpha_i+\beta_i}\Theta\left(\frac{\alpha_i}{\alpha_i+\beta_i}-\epsilon\right)\notag\\&+&\frac{1-\alpha_i-\beta_i}{N}\sum_{j}\frac{\beta_j}{\alpha_j+\beta_j}\Theta\left(\frac{\alpha_j}{\alpha_j+\beta_j}-{\epsilon}\right)
\end{eqnarray}
where $\rho$ is the fraction empty nodes, $\sum_i \left(1-\alpha_i(s)-\beta_i(s)\right)=\rho N$.

To disentangle the evolution of $\alpha_i$ and $\beta_i$ we introduce a new set of variables, the {\em node emptiness}, $\gamma_i=1-\alpha_i-\beta_i$, and the {\em difference of fractions} of $A$ and $B$, $\zeta_i=\alpha_i-\beta_i$.  The new variables range in $\zeta_i\in[-1,1]$ and $\gamma_i\in[0,1]$. Introducing the functions
\begin{eqnarray*}
\label{eq:FG}
F(x)&=&x\Theta(1-x-\epsilon)+(1-x)\Theta(x-\epsilon)\\
G(x)&=&x\Theta(1-x-\epsilon)-(1-x)\Theta(x-\epsilon)\, ,
\end{eqnarray*}
we can rewrite Eqs.~\eqref{eq:alphat1} and~\eqref{eq:betat1} as follows
\begin{equation}
\label{eq:gammat}
\frac{d\gamma_i(s)}{ds}=\rho F\left(\frac{1}{2}+\frac{\zeta_i}{2(1-\gamma_i)}\right)-\frac{\gamma_i}{N}\sum_jF\left(\frac{1}{2}+\frac{\zeta_j}{2(1-\gamma_j)}\right)
\end{equation}
and
\begin{equation}
\label{eq:zetat}
\frac{d\zeta_i(s)}{ds}=-\rho G\left(\frac{1}{2}+\frac{\zeta_i}{2(1-\gamma_i)}\right)+\frac{\gamma_i}{N}\sum_jG\left(\frac{1}{2}+\frac{\zeta_j}{2(1-\gamma_j)}\right)\, .
\end{equation}

\begin{figure}[h]
\begin{center}
\includegraphics[width=9cm]{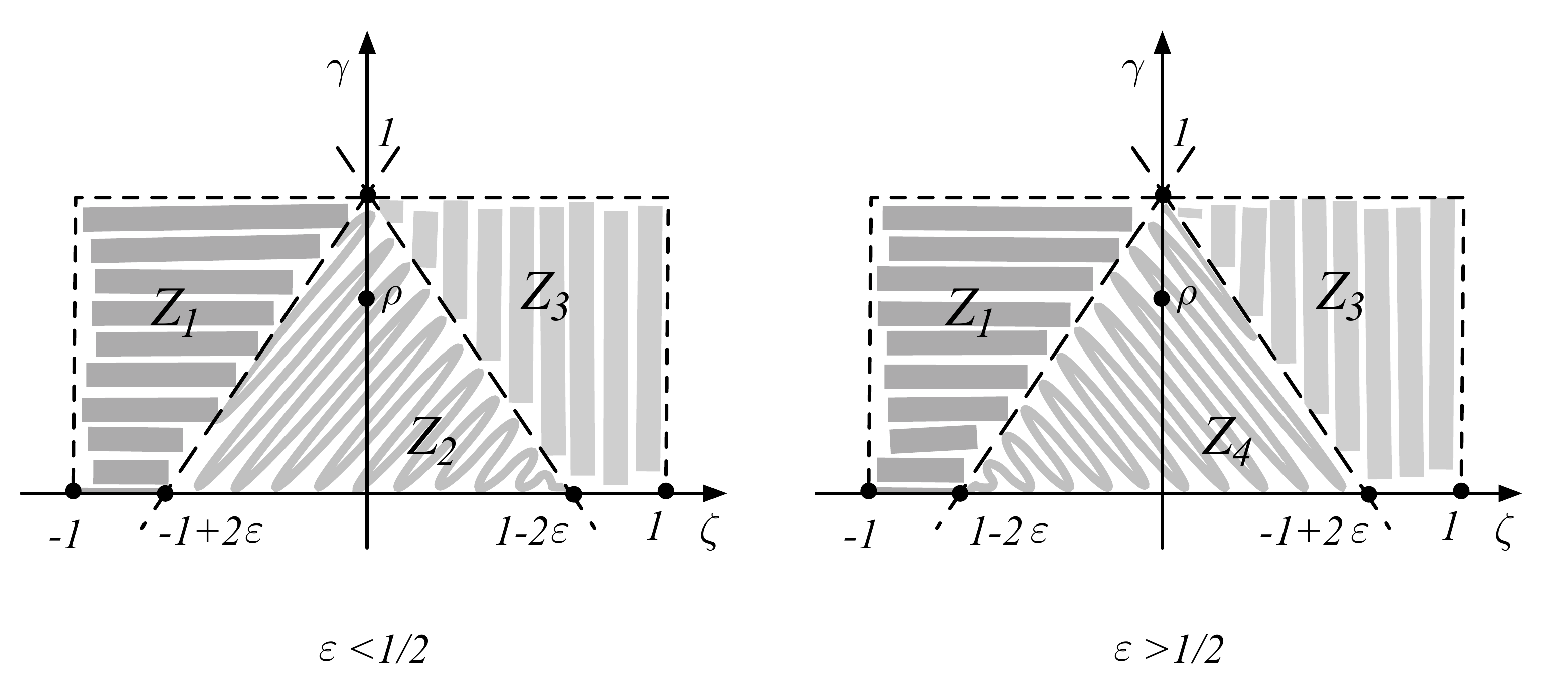}
\end{center}
\caption{The four zones $Z_i$ used to study the solutions of system~\eqref{eq:gammat},~\eqref{eq:zetat}.}
\label{fig:diagramma}
\end{figure}

The agents initialization used in the previous section, namely $\rho N$ density of vacancies and an equal density $(1-\rho)N/2$ of $A$ and $B$ agents, translate into $(\zeta_i,\gamma_i)$ uniformly distributed in a neighborhood of $(0,\rho)$. We thus divide the domain of definition of $(\zeta_i,\gamma_i)$ into four zones (see Fig.~\ref{fig:diagramma}) and we will look closely to the dynamics in the $Z_2$ zone if $\epsilon<1/2$ and in the $Z_4$ zone if $\epsilon>1/2$.

Let $\epsilon<1/2$, then one can easily prove that if $(\zeta,\gamma)\in Z_2$ one has $F\left(\frac{1}{2}+\frac{\zeta}{2(1-\gamma)}\right)=1$ and $G\left(\frac{1}{2}+\frac{\zeta}{2(1-\gamma)}\right)=\frac{\zeta}{1-\gamma}$, so assuming that for all $i$ one has $(\zeta_i(0),\gamma_i(0))\in Z_2$, then Eqs.~\eqref{eq:gammat} and~\eqref{eq:zetat} rewrite:
\begin{equation}
\label{eq:zetagammatZ2}
\begin{cases}
\frac{d\gamma_i(s)}{ds}=\rho -\gamma_i\\
\frac{d\zeta_i(s)}{ds}=-\rho \frac{\zeta_i}{1-\gamma_i}+\frac{\gamma_i}{N}\sum_j \frac{\zeta_j}{1-\gamma_j}\, ,
\end{cases}
\end{equation}
as long as $(\zeta_i(t),\gamma_i(t))$ will not leave $Z_2$. Assume for a while this statement to hold, then the first equation can be straightforwardly solved to give $\gamma_i(t)=\rho+e^{-t}(\gamma_i(0)-\rho)$, that is for all $i$, $\gamma_i(t)\rightarrow \rho$ when $t\rightarrow\infty$. The second equation can also be solved (see Eq. [S13]) and thus to prove that $\zeta_i(t)\rightarrow 0$ for all $i$ when $t\rightarrow 0$. This proves also {\em a posteriori} that $(\zeta_i(t),\gamma_i(t))$ will never leave $Z_2$.

The average magnetization rewrites in such variables as:
\begin{equation}
\label{eq:magnet}
\langle\mu \rangle=\frac{1}{N}\sum_i\frac{\lvert \zeta_i\rvert}{1-\gamma_i}\, ,
\end{equation}
we have hence proved that for initial conditions in $Z_2$ the magnetization asymptotically vanishes (see Fig.~\ref{fig:Mu}).

The remaining case, $\epsilon>1/2$, can be handle as well but it is more cumbersome (see Appendix\ref{appC}). Let us only observe there that for $(\zeta,\gamma)\in Z_4$ one has $F\left(\frac{1}{2}+\frac{\zeta}{2(1-\gamma)}\right)=G\left(\frac{1}{2}+\frac{\zeta}{2(1-\gamma)}\right)=0$, so assuming that for all $i$ one has $(\zeta_i(0),\gamma_i(0))\in Z_4$, then the right hand side of Eqs.~\eqref{eq:gammat} and~\eqref{eq:zetat} identically vanishes and thus the average magnetization depends on the domain where initial conditions have been set. However for a fixed size of the latter, one cannot satisfy the hypothesis $(\zeta_i(0),\gamma_i(0))\in Z_4$ if $\epsilon$ is closer enough but larger then $0.5$, indeed the $Z_4$ zone shrinks to zero in this case (see Fig.~\ref{fig:diagramma}). In this case one should take into account the dynamics of orbits whose initial conditions are set in $Z_1$ and $Z_3$ (see Appendix\ref{sec:analytical} and Figs.~8 and 9). In conclusion the analytical model perfectly fits with the ABM for $\epsilon\geq 0.5$ while it has a different behaviour for $\epsilon<0.5$, because in this range the ABM dynamics is strongly dictated by the stochasticity of the model, orbits tending to converge toward the equilibrium $(0,\rho)$ ($\langle \mu\rangle\sim 0$) are destabilized by fluctuations and thus sent into the zones $Z_1$ and $Z_3$ ($\langle \mu\rangle\sim 1$).

\begin{figure}[h]
\begin{center}
\includegraphics[width=8cm]{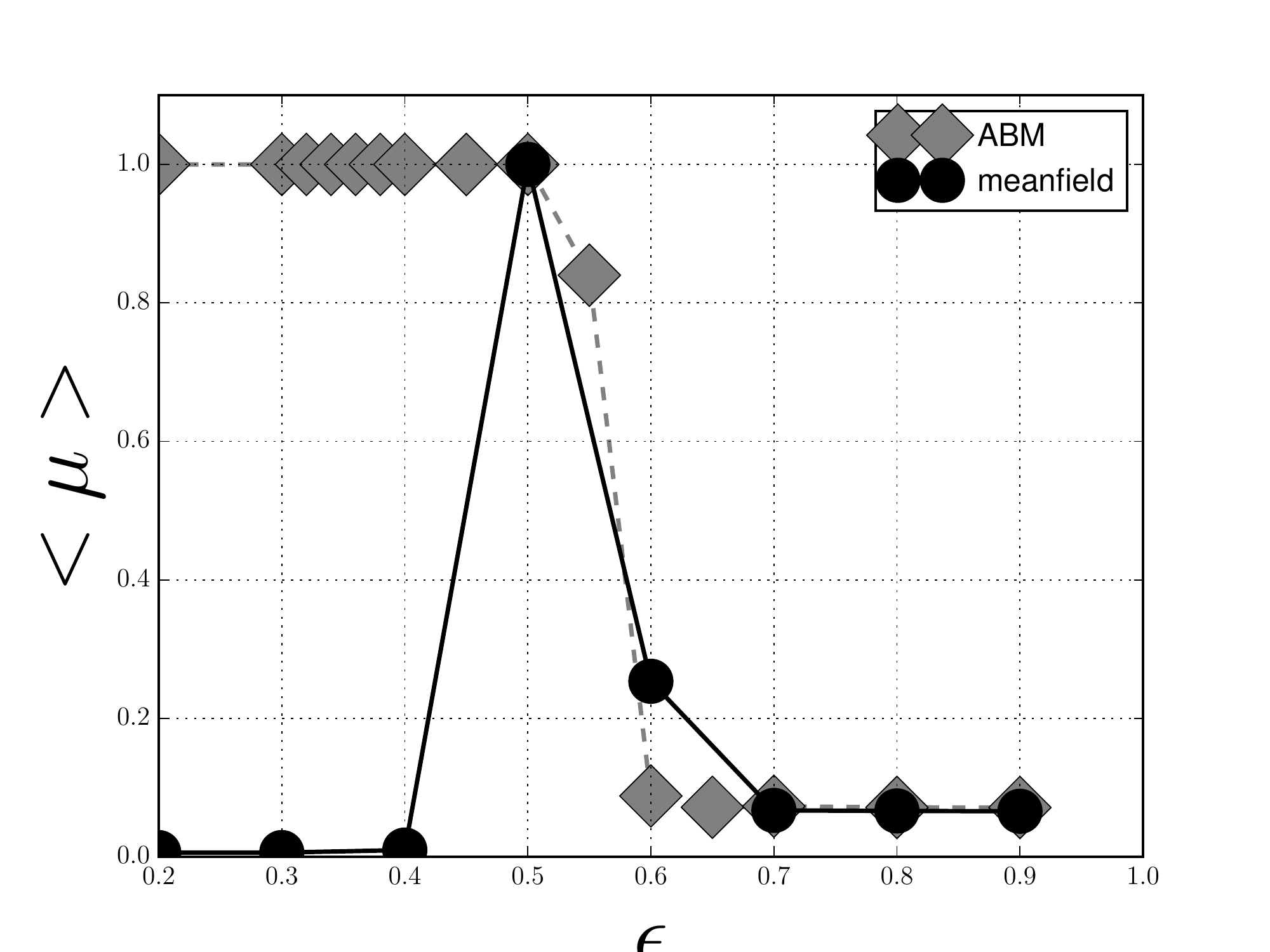}
\end{center}
\caption{The average asymptotic magnetization as a function of $\epsilon$. Each point is the average over $50$ simulations whose initial conditions are close to the equilibrium point $\gamma_i=\rho$ and $\zeta_i=0$. Black circles represent the results of the analytical model while grey diamonds to the ABM with $1$--node fitness. Parameters are: $N=100$ and $\rho=0.9$.}
\label{fig:Mu}
\end{figure}

\section{Conclusions}
\begin{figure*}[h]
\centering
\includegraphics[width=0.8\textwidth]{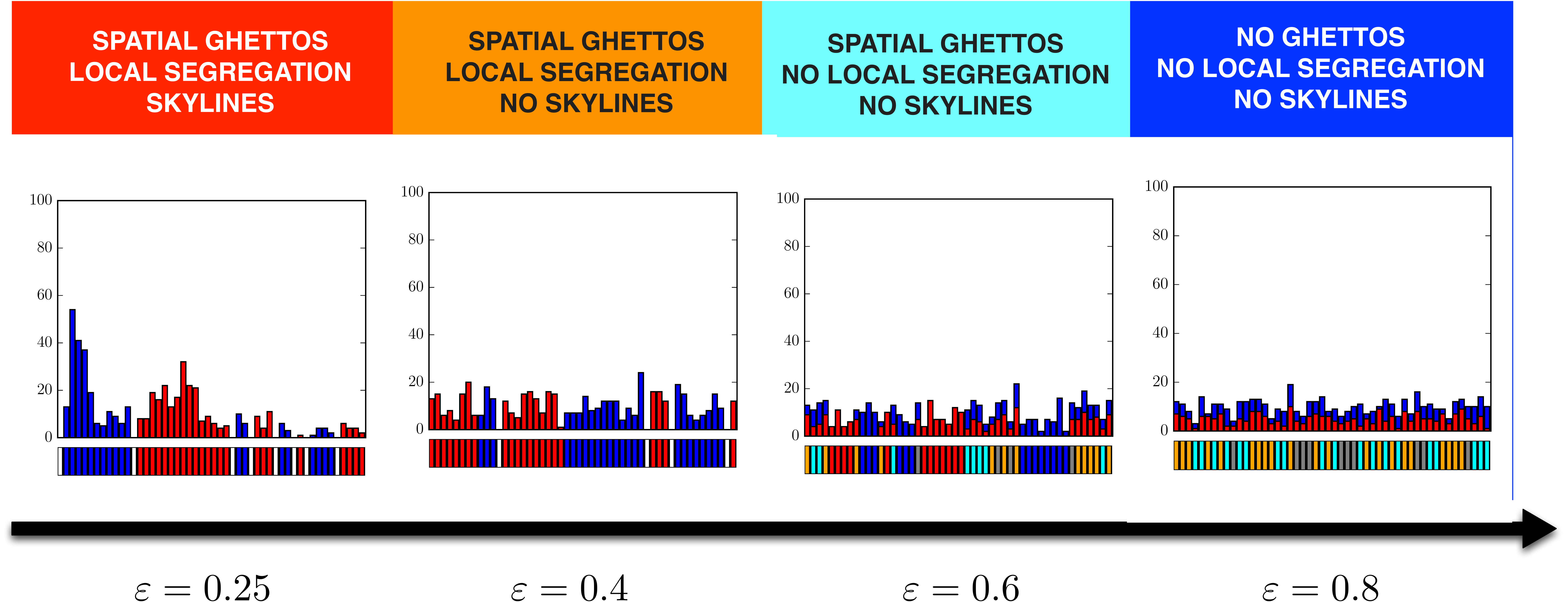}
\caption{\label{regimesSchelling}Summary of the global outcomes of the Schelling metapopulation model on a 1D lattice.  Each box represents a node of the $1$-dimensional lattice; the height of the box is given by the number of Blue and Red agent inside the node. The colored rectangle below the nodes represents the composition of the node obtained using the same color code used in Fig.~\ref{global} Panels A1--A5. }
\end{figure*}

We proposed and analyzed an extension of the classical Schelling model to a metapopulation framework, whose main outcome is the spontaneous emergence, for low values of the tolerance threshold, of heterogeneously populated nodes without any exogenous preferential attachment mechanism (for 1D lattice this phenomenon induces the formation of urban skylines, Fig.~\ref{regimesSchelling}). This behavior is connected to the permanence for long time of the system in a quasi-stationary non segregated state, where in each node the two populations are equally distributed. This quasi-stationary state can be recovered as stable equilibrium of the simplified analytical model. We can evince that the system stabilization toward the magnetized state (for the ABM) passes through the creation (by random agents moves) of highly populated nodes, hereby named \emph{towers}. 
At the same time, global patterns emerge as in the classical Shelling model. Figure~\ref{regimesSchelling} summarizes the possible behaviors of the system. For $\epsilon<0.5$ the global clusters are described as neighborhoods formed by a strong majority of individuals of the same color (soft segregation). For $\epsilon\geq 0.5$ we have the formation of ghettos (hard segregation). \\
The mechanism of formation of the clusters is a typical coarsening phenomenon for low values of $\epsilon$. On the contrary for low tolerance cases the clusters are formed around the towers that become stable points for a certain type of nodes (once an agent, whose kind corresponds to the majority already inside the tower, enters she never gets out). Once a higher density zone starts to exist, this mechanism reinforces the (majority color) population growth in this node and in the neighborhood. The global patterns start therefore to stabilize around the towers.\\
Local magnetization phenomenon has been explained using an analytical approach able to describe the different equilibria of the model and the origin of the quasi-stationary state for low tolerances. 

\section*{Acknowledgments}
The work of F.G., Y.G. and T.C. presents research results of the Belgian
Network DYSCO (Dynamical Systems, Control, and Optimization), funded by
the Interuniversity Attraction Poles Programme, initiated by the Belgian
State, Science Policy Office.\\
T.C. is grateful to Cyril Vargas, student at the ENSEEIHT - INP Toulouse
(France), involved in a preliminary study of this subject during his
Master Degree.

\appendix
\section{Patterns dynamics}
\label{sec:convergence}

The aim of this section is to present some details concerning the convergence of the agent based model to the asymptotic equilibrium pattern. We start by considering the {\em convergence time}, defined as the time needed for the system to reach such equilibrium where no agents has incentive to move anymore, and its dependence of the tolerance threshold $\epsilon$. Intuitively, if $\epsilon$ is large then most agents are happy and thus they will not move, hence the equilibrium will be reached quite soon. On the other hand if $\epsilon$ is small, agents will be very often unhappy and thus they will relocate themselves to reduce this uneasiness, this will increase the convergence time. In the limit $\epsilon\rightarrow 0$ this equilibrium will be never achieved and thus the convergence time will diverge.

\begin{figure}[h]
\centering
\includegraphics[width=0.6\textwidth]{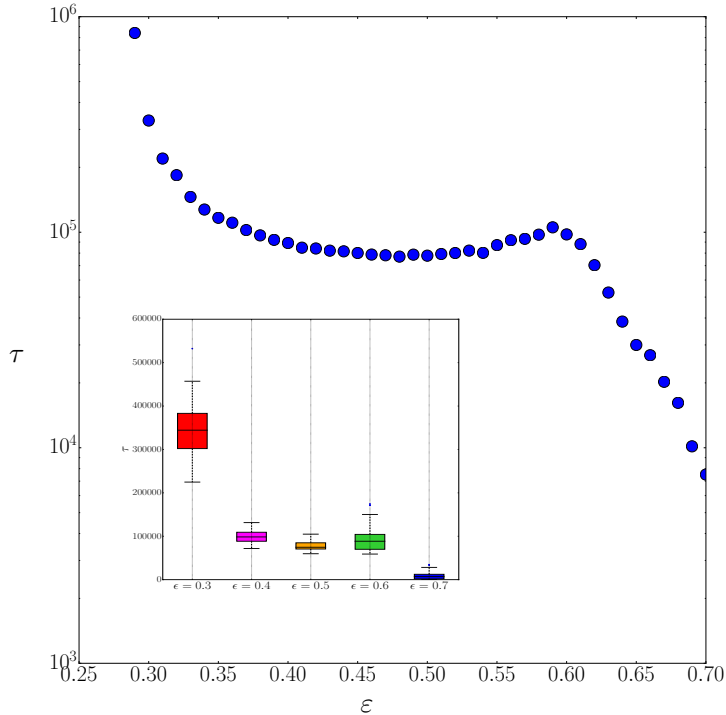}
\caption{\label{convTimes} Convergence time as a function of $\epsilon$. The blue line represents the average convergence time as a function of $\epsilon$ for $100$ replicas of the system with same initial conditions and parameters values. The inset shows the boxplots for the convergence times for some representative values of $\epsilon=0.3,0.4,0.5,0.6,0.7$.}
\end{figure}

In Fig.~\ref{convTimes} we report the results obtained by the simulation of the metapopulation model and we can observe that the convergence time does not have a monotonic behavior as a function of $\epsilon$. Passing from from $\epsilon=1$ to $\epsilon\sim 0.6$ we observe an exponential growth of the convergence time (note the logarithmic vertical scale), these values of $\epsilon$ correspond to the absence of macroscopic clusters (see Fig. 5 main text). Let us notice that the time to converge increase between $\epsilon\sim 0.5$ and $\epsilon\sim 0.6$ is due to a slower dynamics for the latter case (to move is less probable). Then for $0.3\leq\epsilon< 0.6$ the convergence time exhibits an almost stable value - horizontal plateau - with a local minimum at $\epsilon=0.5$, this is the phase where macroscopic clusters are formed through a coarsening process (see Fig. 5 main text). Finally for $\epsilon\leq 0.3$ the convergence times start to increase faster than exponentially, the dynamics exhibits quasi-stationary states and the system reaches the equilibrium through the formation of towers (high densely populated set of contiguous nodes). Let us observe that asymptotic equilibrium is affected by the intrinsic stochasticity of the ABM, the convergence time will thus reflect this fact by showing a large variance, mainly for small $\epsilon$ (see inset of Fig.~\ref{convTimes}).

\begin{figure}[ht]
\centering
\includegraphics[width=1\textwidth]{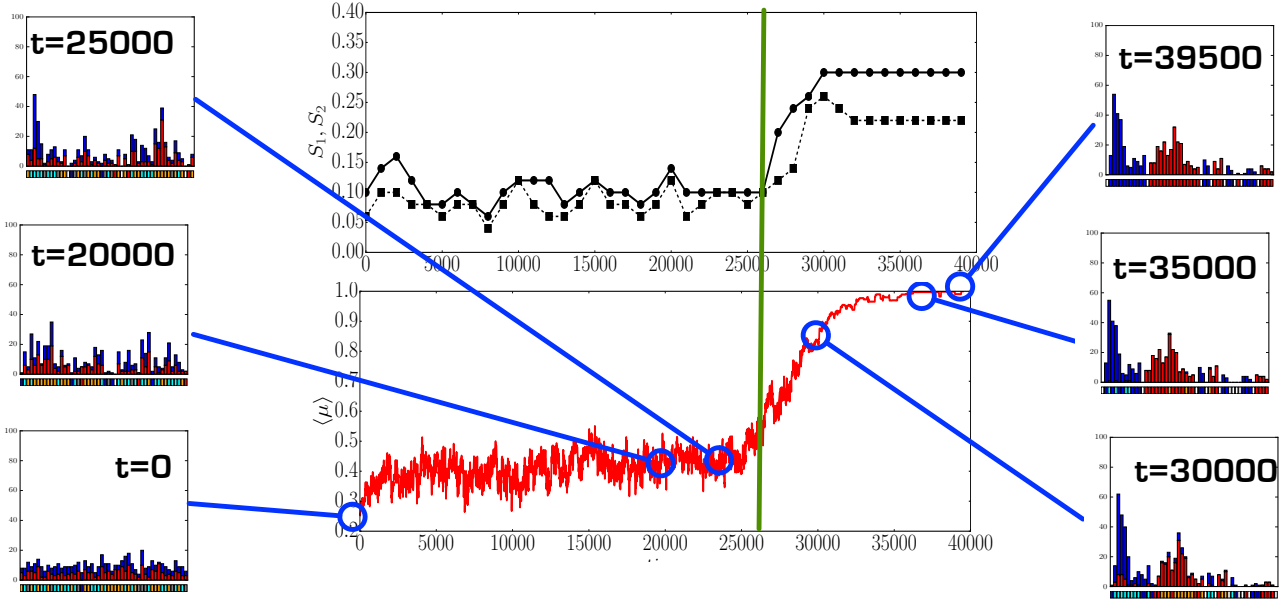}
\caption{\label{timeEvol}Generic time evolution toward a hard segregated pattern. Upper main plot: first and second cluster size as a function of time. Lower main plot: average magnetization of the system as a function of time. Left and right panels: System snapshots at different times. Each box represents a node of the $1$-dimensional lattice, the height of the box is given by the total blue and red population of the node. The colored rectangle under the cell represents the majority color of the cell.}
\end{figure}

The patterns arising for small tolerance threshold are intriguing and peculiar to our model. As already observed the system spends quite a long time in a quasi-stationary (almost) homogeneous state and then suddenly jumps to a macroscopic segregated one. Such behavior is schematically represented in Fig.~\ref{timeEvol} in the case of the $1D$-lattice for $\epsilon=0.25$. The system stabilization toward the magnetized state passes through the creation (by random agents moves) of a highly populated nodes, hereby named \emph{towers} corresponding to monochromatic clusters in the $2$D model presented before. Such towers become attracting selective places for each kind of agents: once an agent, whose kind corresponds to the one of the majority already inside the tower, enters she never gets out from the tower because she will be happy there and this will increase the unhappiness of agents of the opposite kind still inside the tower. The net result is that once a highly dense zone starts to exist, this mechanism reinforces the (majority color) population growth in this node and in the neighborhood, because of the way the fitness is computed, using nodes at distance $1$. The global patterns start therefore to stabilize around the towers. As the first kind of agents stabilizes, with a certain delay also the stabilization of the second population is reached. Once the towers/clusters are formed, the complete equilibrium is reached as the interface between Blue and Red zones becomes empty. The lower is the tolerance value $\epsilon$, the higher is the initial density unbalance needed to start the stabilization process.

\section{More details about the analytical model}
\label{sec:analytical}

The goal of this section is to present more details of the analytical model we introduced to capture the main behavior of the metapopulation Schelling model presented in the main text.

Let us denote by $n^A_i(t)$, respectively $n^B_i(t)$, the number of agents of type $A$, respectively $B$, at node $i$--th at time $t$. Each node is also characterized by a number of vacancies $n^E_i(t)$ at time $t$, however for all $t$ and all $i$ one has $n^A_i(t)+n^B_i(t)+n^E_i(t)=L$ and moreover the number of agents of each kind is a preserved quantity, i.e. the system is closed. The state of the system is thus completely characterized by knowledge of $(\vec{n}^A(t),\vec{n}^B(t))$, being $\vec{n}^A(t)=(n^A_1(t),\dots,n^A_N(t))$ and $\vec{n}^B(t)=(n^B_1(t),\dots,n^B_N(t))$.

An agent $A$ in node $i$--th is unhappy - or her fitness is low - if the fraction of $B$ in the same node, that is $n^B_i/(n^A_i+n^B_i)$, is larger than a given {\em tolerance threshold} $\epsilon\in(0,1)$:
\begin{equation}
\label{eq:Aunhappy}
\text{$A$ is unhappy at $i$ if }\frac{n^B_i}{n^A_i+n^B_i}\geq \epsilon\, ,
\end{equation}
and similarly for $B$.

We assume that once an agent decides to move because unhappy, she will move to an uniformly randomly chosen new node, provide there is enough space there (that we named {\em weak liquid Schelling model}). The model is intrinsically stochastic and hence it can be described by the probability to be at time $t$ in state $(\vec{n}^A,\vec{n}^B)$, that is $P(\vec{n}^A,\vec{n}^B,t)$. The evolution of such probability can be obtained using the master equation:
\begin{eqnarray}
\label{eq:ME}
P(\vec{n}^A,\vec{n}^B,t+1)=P(\vec{n}^A,\vec{n}^B,t)+\sum_{(\vec{n}^{A^\prime},\vec{n}^{B^\prime})} \left[ T(\vec{n}^A,\vec{n}^B\vert\vec{n}^{A^\prime},\vec{n}^{B^\prime})P(\vec{n}^{A^\prime},\vec{n}^{B^\prime},t)-T(\vec{n}^{A^\prime},\vec{n}^{B^\prime}\vert\vec{n}^A,\vec{n}^B)P(\vec{n}^A,\vec{n}^B,t)\right]\, ,
\end{eqnarray}
being $T(\vec{n}^{A^\prime},\vec{n}^{B^\prime}\vert\vec{n}^A,\vec{n}^B)$ the {\em Transition probability} to pass from state $(\vec{n}^A,\vec{n}^B)$ to the new compatible one, i.e. the system can pass from the former to the latter, $(\vec{n}^{A^\prime},\vec{n}^{B^\prime})$. Non compatible states cannot be linked and thus we will set $T(\vec{n}^{A^\prime},\vec{n}^{B^\prime}\vert\vec{n}^A,\vec{n}^B)=0$.

The non zero transition probabilities can be computed using the behavioral rules of the model:
\begin{eqnarray}
\label{eq:transprob}
\text{$A$ moves from node $i$ to node $j$: }T_1(n^A_i-1,n^A_j+1\vert n^A_i,n^A_j)&=& \frac{1}{N}\frac{n^A_i}{n^A_i+n^B_i}\Theta\left(\frac{n^B_i}{n^A_i+n^B_i}-\epsilon\right)\frac{n^E_j}{L}\notag\\
\text{$A$ moves from node $j$ to node $i$: }T_2(n^A_i+1,n^A_j-1\vert n^A_i,n^A_j)&=& \frac{1}{N}\frac{n^A_j}{n^A_j+n^B_j}\Theta\left(\frac{n^B_j}{n^A_j+n^B_j}-\epsilon\right)\frac{n^E_i}{L}\notag\\
\text{$B$ moves from node $i$ to node $j$: }T_3(n^B_i-1,n^B_j+1\vert n^B_i,n^B_j)&=& \frac{1}{N}\frac{n^B_i}{n^A_i+n^B_i}\Theta\left(\frac{n^A_i}{n^A_i+n^B_i}-\epsilon\right)\frac{n^E_j}{L}\notag\\
\text{$B$ moves from node $j$ to node $i$: }T_4(n^B_i+1,n^B_j-1\vert n^B_i,n^B_j)&=& \frac{1}{N}\frac{n^B_j}{n^A_j+n^B_j}\Theta\left(\frac{n^A_j}{n^A_j+n^B_j}-\epsilon\right)\frac{n^E_i}{L}\, .
\end{eqnarray}
 Let us observe that to lighten the notation we wrote only the variables whose values change because of the transition. The function $\Theta(x)$ is defined to be $1$ if $x>0$ and $0$ otherwise; smoother versions can be used as well. Being the structure of such transition probabilities very similar we will detail the way we compute the first one:
\begin{itemize}
\item $\frac{1}{N}$ : probability to draw the $i$--th node with uniform random probability;
\item $\frac{n^A_i}{n^A_i+n^B_i}$ : probability to draw one $A$ agent among the $n^A_i$ present over the total node population $n^A_i+n^B_i$;
\item $\Theta\left(\frac{n^B_i}{n^A_i+n^B_i}-\epsilon\right)$: probability the selected $A$ agent is unhappy;
\item $\frac{n^E_j}{L}$: probability to select a vacancy in node $j$.
\end{itemize}

The master equation is unmanageable but one can go one step further by computing the time evolution of relevant quantities, for instance the average number of agents $A$ and $B$ in every node $i$ at time $t$, $\langle n^A_i\rangle(t)=\sum_{\vec{n}^A}n^A_iP(\vec{n}^A,\vec{n}^B,t)$:
\begin{eqnarray*}
\langle n^A_i\rangle(t+1)&-&\langle n^A_i\rangle(t)=\sum_{\vec{n}^A}\sum_{j\neq i} \Big[n^A_iT_1(n^A_i,n^A_j\vert n^A_i+1,n^A_j-1)P(n^A_i+1,n^A_j-1,t)\\
&+&n^A_iT_2(n^A_i,n^A_j\vert n^A_i-1,n^A_j+1)P(n^A_i-1,n^A_j+1,t)-n^A_iT_1(n^A_i-1,n^A_j+1\vert n^A_i,n^A_j)P(n^A_i,n^A_j,t)\\&-&n^A_iT_2(n^A_i+1,n^A_j-1\vert n^A_i,n^A_j)P(n^A_i,n^A_j,t)\Big]\\
&=&-\sum_{j\neq i}\langle T_1(n^A_i-1,n^A_j+1\vert n^A_i,n^A_j)\rangle+\sum_{j\neq i}\langle T_2(n^A_i+1,n^A_j-1\vert n^A_i,n^A_j)\rangle\, .
\end{eqnarray*}
And similarly for $B$:
\begin{eqnarray*}
\langle n^B_i\rangle(t+1)-\langle n^B_i\rangle(t)=-\sum_{j\neq i}\langle T_3(n^B_i-1,n^B_j+1\vert n^B_i,n^B_j)\rangle+\sum_{j\neq i}\langle T_4(n^B_i+1,n^B_j-1\vert n^B_i,n^B_j)\rangle\, .
\end{eqnarray*}

Using the expression for the transition probabilities~\eqref{eq:transprob} and assuming correlations can be neglected, i.e. $\langle (n^A_i)^2\rangle\sim \langle n^A_i\rangle^2$, we get:
\begin{eqnarray}
\label{eq:at}
\langle n^A_i\rangle(t+1)-\langle n^A_i\rangle(t)&=&-\frac{1}{N}\sum_{j\neq i}\frac{\langle n^A_i\rangle}{\langle n^A_i\rangle+\langle n^B_i\rangle}\Theta\left(\frac{\langle n^B_i\rangle}{\langle n^A_i\rangle+\langle n^B_i\rangle}-\epsilon\right)\frac{L-\langle n^A_j\rangle-\langle n^B_j\rangle}{L}\notag\\&+&\frac{1}{N}\sum_{j\neq i}\frac{\langle n^A_j\rangle}{\langle n^A_j\rangle+\langle n^B_j\rangle}\Theta\left(\frac{\langle n^B_j\rangle}{\langle n^A_j\rangle+\langle n^B_j\rangle}-\epsilon\right)\frac{L-\langle n^A_i\rangle-\langle n^B_i\rangle}{L}\, .
\end{eqnarray}
and
\begin{eqnarray}
\label{eq:bt}
\langle n^B_i\rangle(t+1)-\langle n^B_i\rangle(t)&=&-\frac{1}{N}\sum_{j\neq i}\frac{\langle n^B_i\rangle}{\langle n^A_i\rangle+\langle n^B_i\rangle}\Theta\left(\frac{\langle n^A_i\rangle}{\langle n^A_i\rangle+\langle n^B_i\rangle}-\epsilon\right)\frac{L-\langle n^A_j\rangle-\langle n^B_j\rangle}{L}\notag\\&+&\frac{1}{N}\sum_{j\neq i}\frac{\langle n^B_j\rangle}{\langle n^A_j\rangle+\langle n^B_j\rangle}\Theta\left(\frac{\langle n^A_j\rangle}{\langle n^A_j\rangle+\langle n^B_j\rangle}-\epsilon\right)\frac{L-\langle n^A_i\rangle-\langle n^B_i\rangle}{L}\, .
\end{eqnarray}
Observe that the right hand sides of~\eqref{eq:at} and~\eqref{eq:bt} remain unchanged if we allow both sums to run over all node indexes, namely include also $j=i$.

To go one step further, let us define the fraction of $A$ and $B$ in each node, that is
\begin{equation}
\label{eq:fracAB}
\alpha_i=\frac{\langle n^A_i\rangle}{L}\text{ and }\beta_i=\frac{\langle n^B_i\rangle}{L}\, ,
\end{equation}
then we can rewrite~\footnote{Let us observe that $\langle n^A_i\rangle/(\langle n^A_i\rangle+\langle n^B_i\rangle)-\epsilon$ is positive if and only if $\alpha_i/(\alpha_i+\beta_i)-{\epsilon}>0$, and similarly for the other term.} the previous equations~\eqref{eq:at} and~\eqref{eq:bt} in terms of $\alpha_i$ and $\beta_i$.
%
Finally rescaling time by $s=t/L$, 
dividing the equations for $\alpha_i$ and $\beta_i$ by $1/L$ and passing to the limit $L\rightarrow +\infty$ we get:
\begin{eqnarray}
\label{eq:alphat1}
\frac{d\alpha_i(s)}{ds}&=&\lim_{L\rightarrow+\infty}\frac{\alpha_i(s+1/L)-\alpha_i(s)}{1/L}\\&=&-\frac{1}{N}\frac{\alpha_i}{\alpha_i+\beta_i}\Theta\left(\frac{\beta_i}{\alpha_i+\beta_i}-\epsilon\right)\sum_{j}(1-\alpha_j-\beta_j)+\frac{1-\alpha_i-\beta_i}{N}\sum_{j}\frac{\alpha_j}{\alpha_j+\beta_j}\Theta\left(\frac{\beta_j}{\alpha_j+\beta_j}-{\epsilon}\right)\notag\, ,
\end{eqnarray}
and
\begin{eqnarray}
\label{eq:betat1}
\frac{d\beta_i(s)}{ds}&=&\lim_{L\rightarrow+\infty}\frac{\beta_i(s+1/L)-\beta_i(s)}{1/L}\\&=&-\frac{1}{N}\frac{\beta_i}{\alpha_i+\beta_i}\Theta\left(\frac{\alpha_i}{\alpha_i+\beta_i}-\epsilon\right)\sum_{j}(1-\alpha_j-\beta_j)+\frac{1-\alpha_i-\beta_i}{N}\sum_{j}\frac{\beta_j}{\alpha_j+\beta_j}\Theta\left(\frac{\alpha_j}{\alpha_j+\beta_j}-{\epsilon}\right)\notag\, .
\end{eqnarray}


\begin{remark}[Some preserved quantities]
Let us observe that the model correctly preserves the total fractions of $A$ and $B$ agents in time and thus the total vacancies $\sum_i \left(1-\alpha_i(s)-\beta_i(s)\right)=\sum_i \left(1-\alpha_i(0)-\beta_i(0)\right)=\rho N$, where $\rho$ is the emptiness defined previously, i.e. the total fraction of vacancies.

To prove this statement is enough to take the time derivative of $\sum_i \left(1-\alpha_i(s)-\beta_i(s)\right)$ and observe that using Eqs.~\eqref{eq:alphat1} and~\eqref{eq:betat1} one gets: 
\begin{equation*}
\frac{d}{ds}\sum_i \left(1-\alpha_i(s)-\beta_i(s)\right)=0\, .
\end{equation*}

One can similarly prove the statement about the total fraction of $A$ and $B$ agents
\end{remark}

So in conclusion the system is ruled by the following system of differential equations:
\begin{equation}
\label{eq:alphabetat}
\begin{cases}
\frac{d\alpha_i(s)}{ds}=-\rho \frac{\alpha_i}{\alpha_i+\beta_i}\Theta\left(\frac{\beta_i}{\alpha_i+\beta_i}-\epsilon\right)+\frac{(1-\alpha_i-\beta_i)}{N}\sum_{j}\frac{\alpha_j}{\alpha_j+\beta_j}\Theta\left(\frac{\beta_j}{\alpha_j+\beta_j}-{\epsilon}\right)\\
\frac{d\beta_i(s)}{ds}=-\rho\frac{\beta_i}{\alpha_i+\beta_i}\Theta\left(\frac{\alpha_i}{\alpha_i+\beta_i}-\epsilon\right)+\frac{(1-\alpha_i-\beta_i)}{N}\sum_{j}\frac{\beta_j}{\alpha_j+\beta_j}\Theta\left(\frac{\alpha_j}{\alpha_j+\beta_j}-{\epsilon}\right)\, .
\end{cases}
\end{equation}

\section{The average magnetization}
\label{appC}
To better analyze the system and in particular be able to describe the dependence of the average magnetization on $\epsilon$, we introduce a new set of variables, the {\em local emptiness}, $\gamma_i=1-\alpha_i-\beta_i$, and the {\em local difference of fractions} of $A$ and $B$, $\zeta_i=\alpha_i-\beta_i$. The original coordinates can be obtained back using $\alpha_i=(1-\gamma_i+\zeta_i)/2$ and  $\beta_i=(1-\gamma_i-\zeta_i)/2$. The new variables ranges are $\zeta_i\in[-1,1]$ and $\gamma_i\in[0,1]$ and the magnetization rewrites as
\begin{equation}
\label{eq:magnezg}
\langle \mu\rangle =\frac{1}{N}\sum_i\frac{|\zeta_i|}{1-\gamma_i}\, .
\end{equation}

Let us introduce the functions
\begin{equation}
\label{eq:FG}
F(x):=x\Theta(1-x-\epsilon)+(1-x)\Theta(x-\epsilon)\text{ and }G(x):=x\Theta(1-x-\epsilon)-(1-x)\Theta(x-\epsilon)\, ,
\end{equation}
then observing that
\begin{equation*}
\frac{\alpha_i}{\alpha_i+\beta_i}=\frac{1}{2}+\frac{\zeta_i}{2(1-\gamma_i)} \text{ and }\frac{\beta_i}{\alpha_i+\beta_i}=\frac{1}{2}-\frac{\zeta_i}{2(1-\gamma_i)}\, ,
\end{equation*}
we can rewrite Eq.~\eqref{eq:alphabetat} as follows
\begin{equation}
\label{eq:zetagammat}
\begin{cases}
\frac{d\gamma_i(s)}{ds}=\rho F\left(\frac{1}{2}+\frac{\zeta_i}{2(1-\gamma_i)}\right)-\frac{\gamma_i}{N}\sum_jF\left(\frac{1}{2}+\frac{\zeta_j}{2(1-\gamma_j)}\right)\\
\frac{d\zeta_i(s)}{ds}=-\rho G\left(\frac{1}{2}+\frac{\zeta_i}{2(1-\gamma_i)}\right)+\frac{\gamma_i}{N}\sum_jG\left(\frac{1}{2}+\frac{\zeta_j}{2(1-\gamma_j)}\right)\, .
\end{cases}
\end{equation}

To qualitatively study the solutions of the latter system we define four zones in the $(\zeta,\gamma)$ plane as follows (see Fig. 3 main text):
\begin{eqnarray*}
Z_1&=&\{(\zeta,\gamma)\in[-1,1]\times [0,1]: \frac{1}{2}-\frac{\zeta_i}{2(1-\gamma_i)}-\epsilon \geq 0 \text{ and } \frac{1}{2}+\frac{\zeta_i}{2(1-\gamma_i)}-\epsilon < 0\}\\
Z_2&=&\{(\zeta,\gamma)\in[-1,1]\times [0,1]: \frac{1}{2}-\frac{\zeta_i}{2(1-\gamma_i)}-\epsilon \geq 0 \text{ and } \frac{1}{2}+\frac{\zeta_i}{2(1-\gamma_i)}-\epsilon \geq 0\}\\
Z_3&=&\{(\zeta,\gamma)\in[-1,1]\times [0,1]: \frac{1}{2}-\frac{\zeta_i}{2(1-\gamma_i)}-\epsilon < 0 \text{ and } \frac{1}{2}+\frac{\zeta_i}{2(1-\gamma_i)}-\epsilon \geq 0\}\\
Z_4&=&\{(\zeta,\gamma)\in[-1,1]\times [0,1]: \frac{1}{2}-\frac{\zeta_i}{2(1-\gamma_i)}-\epsilon < 0 \text{ and } \frac{1}{2}+\frac{\zeta_i}{2(1-\gamma_i)}-\epsilon < 0\}\, .
\end{eqnarray*}
Let us observe that if $\epsilon<1/2$ then $Z_4$ is empty and if $\epsilon>1/2$ then $Z_2$ is empty.

The initial condition of the Shelling model presented in the main text translate into $(\zeta_i,\gamma_i)$ distributed close to $(0,\rho)$. We will thus be interested in the behavior of the solutions of Eq.~\eqref{eq:zetagammat} in the $Z_2$ zone if $\epsilon<1/2$ and in the $Z_4$ zone if $\epsilon>1/2$.

Let us consider the case $\epsilon<1/2$, then one can easily prove that for all $(\zeta,\gamma)\in Z_2$ one has $F\left(\frac{1}{2}+\frac{\zeta}{2(1-\gamma)}\right)=1$ and $G\left(\frac{1}{2}+\frac{\zeta}{2(1-\gamma)}\right)=\frac{\zeta}{1-\gamma}$, so assuming that for all $i$ one has $(\zeta_i(0),\gamma_i(0))\in Z_2$, then:
\begin{equation}
\label{eq:zetagammatZ2}
\begin{cases}
\frac{d\gamma_i(s)}{ds}=\rho -\gamma_i\\
\frac{d\zeta_i(s)}{ds}=-\rho \frac{\zeta_i}{1-\gamma_i}+\frac{\gamma_i}{N}\sum_j \frac{\zeta_j}{1-\gamma_j}\, ,
\end{cases}
\end{equation}
as long as $(\zeta_i(t),\gamma_i(t))$ will not leave $Z_2$. Assume for a while this statement, then the first equation can be straightforwardly solved to give $\gamma_i(t)=\rho+e^{-t}(\gamma_i(0)-\rho)$, that is for all $i$, $\gamma_i(t)\rightarrow \rho$ when $t\rightarrow\infty$.

The solution for the second is more cumbersome, but one prove the existence of a lower (upper) solution $\zeta_i^-(t)$ ($\zeta_i^+(t)$) such that $\zeta_i^-(t)<\zeta_i(t)<\zeta_i^+(t)$ where:
\begin{equation}
\label{eq:explsol}
\zeta^{\pm}_i(t)=\left(\frac{e^{-t}(1-\gamma_i(0))}{1-\rho-e^{-t}(\gamma_i(0)-\rho)}\right)^{\frac{\rho}{1-\rho}}\left[\zeta_i(0)\mp(1-2\epsilon)\left(\frac{\gamma_i(0)-1}{1-\gamma_i(0)}\right)^{\frac{\rho}{1-\rho}}\int_{\gamma_i(0)}^{\rho+e^{-t}(\gamma_i(0)-\rho)}\frac{x(1-x)^{\frac{\rho}{1-\rho}}}{(x-\rho)^{\frac{\rho}{1-\rho}}}dx\right]\, ,
\end{equation}
from which one can get $\zeta_i(t)\rightarrow 0$ for all $i$ when $t\rightarrow 0$.

Because the point $(0,\rho)$ belongs to $Z_2$ we have {\em a posteriori proved} that $(\zeta_i(t),\gamma_i(t))$ will never leave $Z_2$. A simpler but also weaker statement can be obtained by observing that the equilibrium $(\zeta_i,\gamma_i)=(0,\rho)$ for all $i$ is stable being its eigenvalues $-\rho/(1-\rho)$ and $-1$, each one with multiplicity $N$, thus there exists a neighborhood of $(0,\rho)$ such that all orbits whose initial conditions are inside never leave it. Using the definition of $\langle \mu\rangle$ given by Eq.~\eqref{eq:magnezg}, we have thus proven that for $\epsilon<0.5$ the average magnetization converges asymptotically to $0$ (see Fig.~4 main text).

Assuming now $\epsilon>1/2$, then one can easily prove that for all $(\zeta,\gamma)\in Z_4$ one has $F\left(\frac{1}{2}+\frac{\zeta}{2(1-\gamma)}\right)=G\left(\frac{1}{2}+\frac{\zeta}{2(1-\gamma)}\right)=0$, so assuming that for all $i$ one has $(\zeta_i(0),\gamma_i(0))\in Z_4$, then:
\begin{equation}
\label{eq:zetagammatZ4}
\begin{cases}
\frac{d\gamma_i(s)}{ds}=0\\
\frac{d\zeta_i(s)}{ds}=0\, ,
\end{cases}
\end{equation}
and thus $(\zeta_i(t),\gamma_i(t))$ will not evolve and thus remain in $Z_4$. However fixed the size of the domain centred at $(\zeta,\gamma)=(0,\rho)$, where initial conditions are taken, the above assumption cannot be satisfied if $\epsilon >0.5$ but close to it. Indeed looking at the right panel of Fig. 3 in the main text, we realize that the zone $Z_4$ shrinks to zero as $\epsilon\rightarrow 0.5$ from above. The aim of the rest of this section is to prove that we can take this fact into account and explain the behavior of $\langle \mu\rangle$ also for $\epsilon >0.5$.

As already remarked the functions $F$ and $G$ vanish for $(\zeta,\gamma)\in Z_4$, thus Eq.~\eqref{eq:zetagammat} can be rewritten as:
\begin{equation}
\label{eq:zetagammat2}
\begin{cases}
\frac{d\gamma_i(s)}{ds}=\rho F\left(\frac{1}{2}+\frac{\zeta_i}{2(1-\gamma_i)}\right)-\frac{\gamma_i}{N}\left[\sum_{j\in Z_3}F\left(\frac{1}{2}+\frac{\zeta_j}{2(1-\gamma_j)}\right)+\sum_{j\in Z_1}F\left(\frac{1}{2}+\frac{\zeta_j}{2(1-\gamma_j)}\right)\right]\\
\frac{d\zeta_i(s)}{ds}=-\rho G\left(\frac{1}{2}+\frac{\zeta_i}{2(1-\gamma_i)}\right)+\frac{\gamma_i}{N}\left[\sum_{j\in Z_3}G\left(\frac{1}{2}+\frac{\zeta_j}{2(1-\gamma_j)}\right)+\sum_{j\in Z_1}G\left(\frac{1}{2}+\frac{\zeta_j}{2(1-\gamma_j)}\right)\right]\, ,
\end{cases}
\end{equation}
where we used the shortened notation $j\in Z_k$, to mean $(\zeta_j,\gamma_j)\in Z_k$ and where the zones $Z_1$ and $Z_3$ have been defined in the Fig.~3 of the main text.

\begin{figure}[h]
\centering
\includegraphics[width=0.45\textwidth]{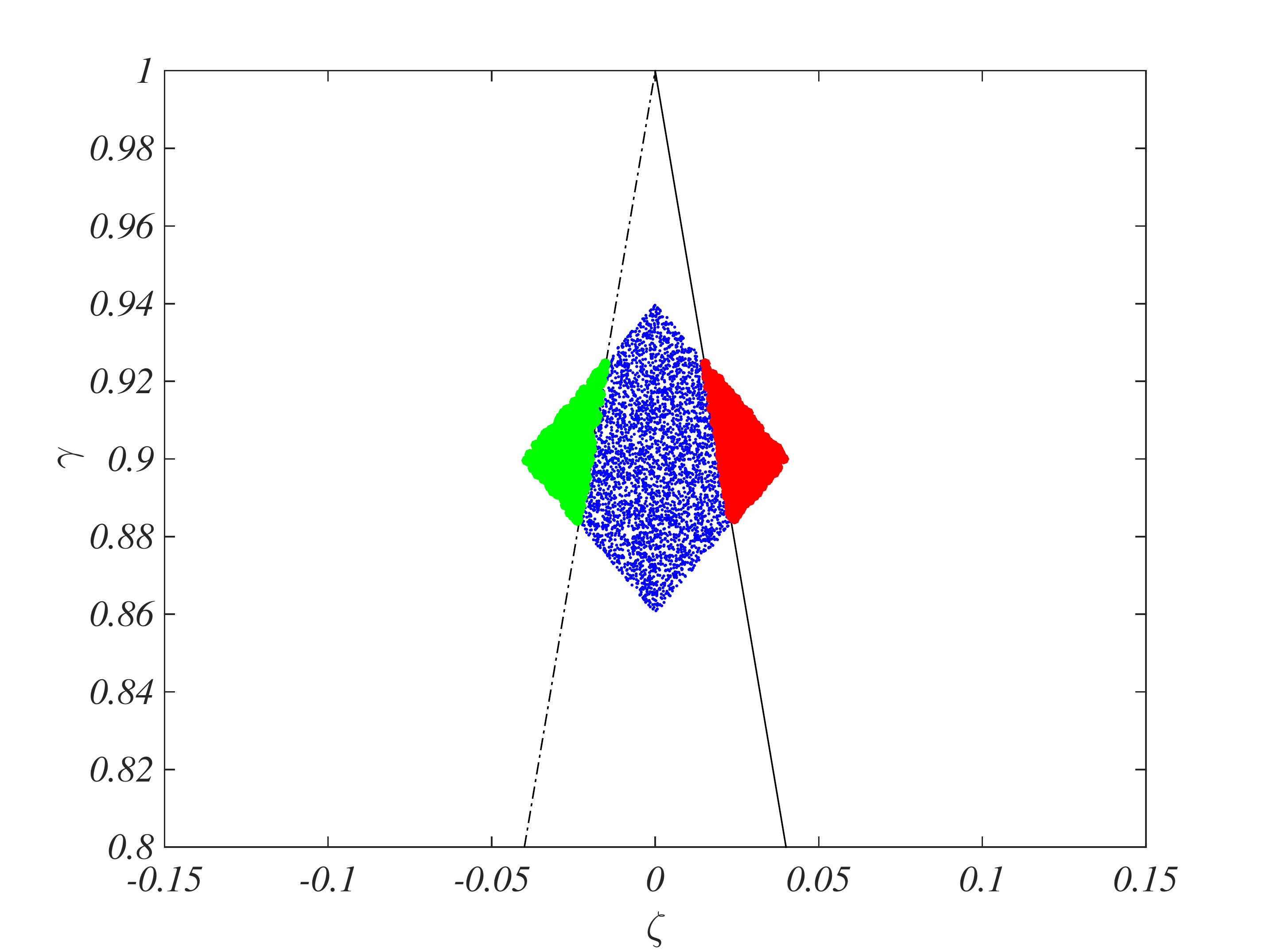}\quad
\includegraphics[width=0.45\textwidth]{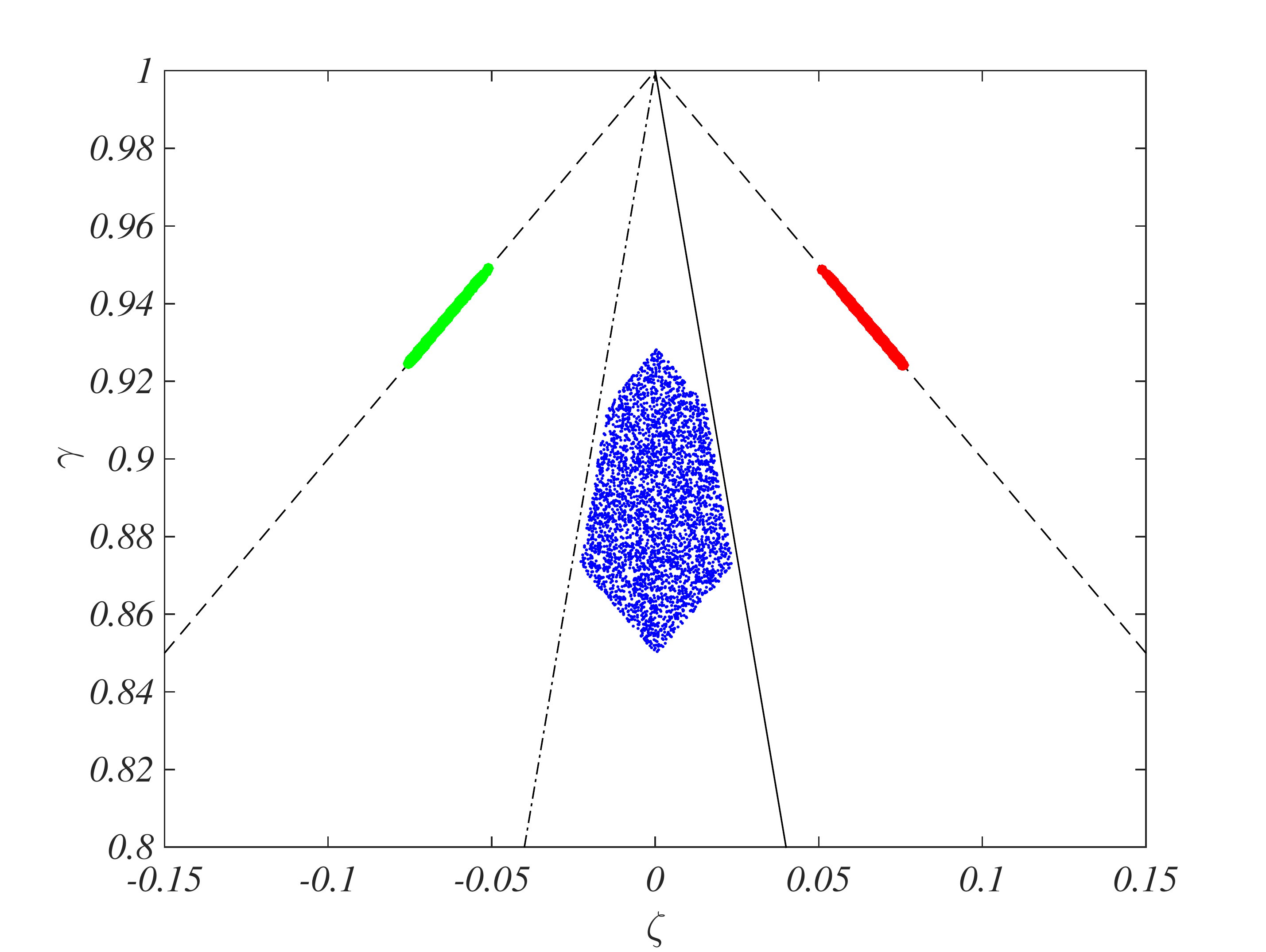}
\caption{\label{dynZ1Z3} Dynamics for $\epsilon>0.5$. Left panel: $5000$ initial conditions are uniformly drawn $(\alpha_i,\beta_i)\in(1-\rho)/2+\delta U[-1,1]$ and then transformed into the new variables $(\zeta_i,\gamma_i)$; points in $Z_1$ are colored in green, points in $Z_4$ in blue and points in $Z_3$ in red. Right panel: asymptotic configuration of the orbits corresponding to the initial conditions given in the left panel, each point is drawn using its initial color. The black solid line is the straight line $\zeta=(2\epsilon -1)(1-\gamma)$, the dot-dashed line is the straight line $\zeta=(1-2\epsilon)(1-\gamma)$ and the dashed lines (right panel) are the straight lines $\zeta=\pm(1-\gamma)$.}
\end{figure}

The initialization of the metapopulation model, translates into original variables $(\alpha_i,\beta_i)$ initially uniformly randomly distributed in $(1-\rho)/2+\delta U[-1,1]$, for some positive small $\delta$. Such domain is distorted passing to the new variables $(zeta_i,\gamma_i)$, in particular it is translated into $(0,\rho)$, rotated by $45^{\circ}$ and expanded by a factor $\sqrt{2}$ in both directions (see left panel of Fig.~\ref{dynZ1Z3}).

Let us observe that $F(0)=G(0)=F(1)=G(1)=0$ and thus Eqs.~\eqref{eq:zetagammat2} admit as equilibrium points $\zeta_i=\pm (1-\gamma_i)$. In conclusion, initial conditions inside $Z_4$ will remain there while orbits originated from initial conditions in $Z_1$ and $Z_3$ will converge somewhere onto the straight lines $\zeta=\pm(1-\gamma)$. In the left panel of Fig.~\ref{dynZ1Z3} we report $5000$ initial conditions built using the above described procedure for $\delta=0.02$, points in $Z_1$ are marked in green, points in $Z_4$ in blue and points in $Z_3$ in red. The Eqs.~\eqref{eq:zetagammat2}  are then numerically solved and the asymptotic positions are drawn in the right panel of Fig.~\ref{dynZ1Z3}, giving to any points the color corresponding to the zone from where is started. One can clearly observe that all blue points remain in the $Z_4$ zone, while green (respectively red) points converge to $\zeta=\gamma-1$ (respectively to $\zeta=1-\gamma$ ).

We are now able to provide a good approximation of the average magnetization. First of all let us rewrite Eq.~\eqref{eq:magnezg} as:
\begin{equation*}
\langle \mu\rangle =\frac{1}{N}\left(\sum_{i\in Z_4}\frac{|\zeta_i|}{1-\gamma_i}+\sum_{i\in Z_1\cup Z_3}\frac{|\zeta_i|}{1-\gamma_i}\right)=\frac{N_4}{N}\frac{1}{N_4}\sum_{i\in Z_4}\frac{|\zeta_i|}{1-\gamma_i}+\frac{N_{13}}{N}\frac{1}{N_{13}}\sum_{i\in Z_1\cup Z_3}\frac{|\zeta_i|}{1-\gamma_i}\, ,
\end{equation*}
where $N_4$ is the number of points in the $Z_4$ zone and $N_{13}$ the total number of points in the $Z_1$ and $Z_3$ zones. Because points in $Z_1$ and $Z_3$ converge to $\zeta=\pm(1-\gamma)$ the right most term in the previous equation is trivially equal to $N_{13}/N$. 

\begin{figure}[h]
\centering
\includegraphics[width=0.8\textwidth]{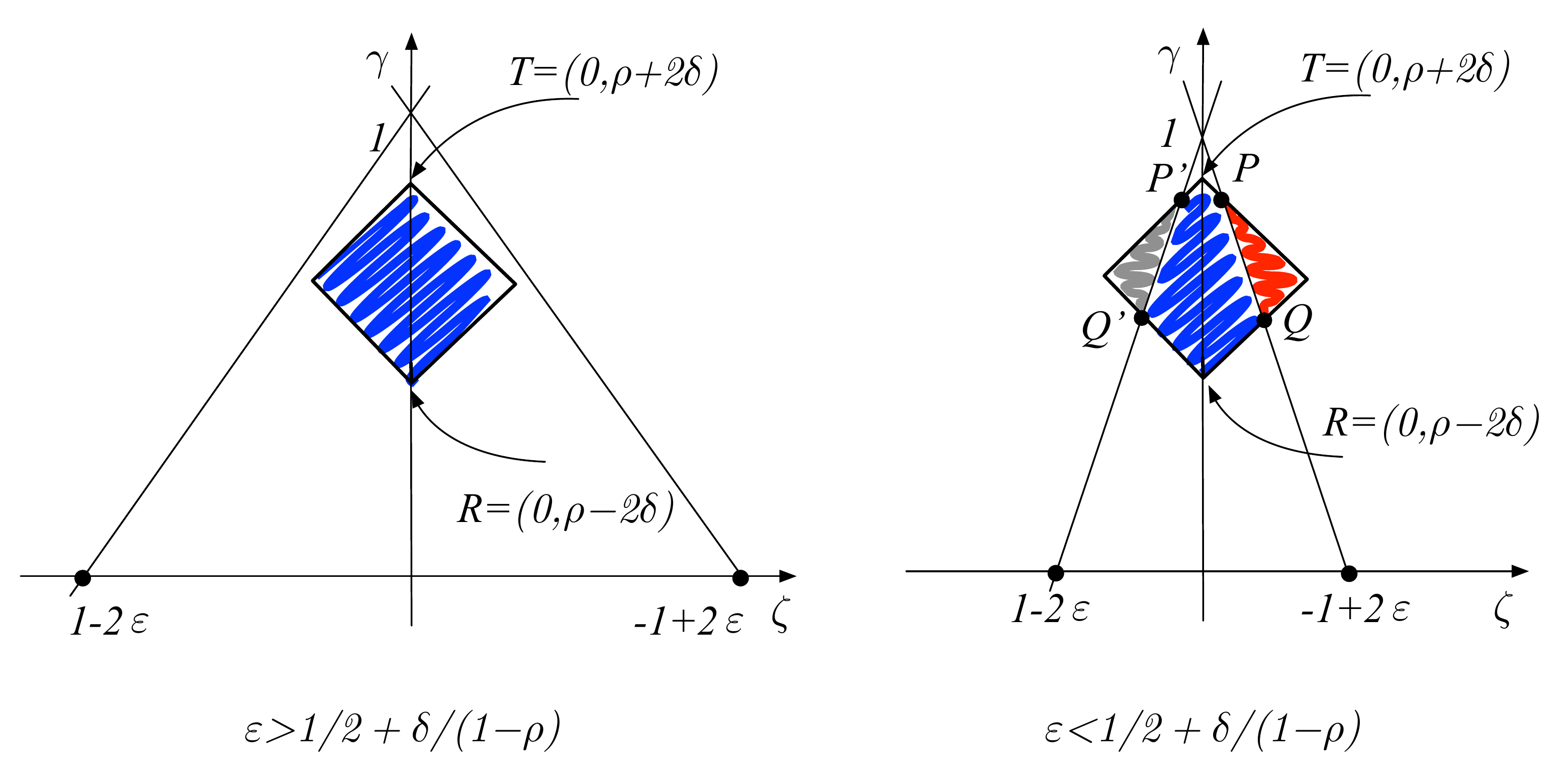}
\caption{\label{domains} The geometries used to compute an approximation to the average magnetization in the case $\epsilon >0.5$.}
\end{figure}

To compute the contribution arising from points in $Z_4$ is more cumbersome, let us hereby stress the main ideas. One can easily show that if $\epsilon >0.5+\delta/(1-\rho)$ then the lines $\zeta=(2\epsilon -1)(1-\gamma)$ and $\zeta=(1-2\epsilon)(1-\gamma)$ do not intersect the diamond like domain (see left panel Fig.~\ref{domains}) and thus all points are initially in the $Z_4$ zones, this means that $N_4=N$ and $N_{13}=0$ so the magnetization has only a contribution from the $Z_4$ zone. On the contrary if $\epsilon <0.5+\delta/(1-\rho)$ some initial conditions are in the $Z_1$ and $Z_3$ zones (see right panel Fig.~\ref{domains}) and thus the magnetization has both contributions.

In the case $\epsilon >0.5+\delta/(1-\rho)$ one can estimate the contribution to the magnetization as:
\begin{equation}
\label{eq:muint1}
\mu_{int,1}=\frac{1}{N_4}\sum_{i\in Z_4}\frac{|\zeta_i|}{1-\gamma_i}\sim\frac{2}{8\delta^2}\int_0^{2\delta}d\zeta\, \zeta\int_{\zeta+\rho-2\delta}^{-\zeta+\rho+2\delta}d\gamma\frac{1}{1-\gamma}=\frac{1}{4\delta^2}\int_0^{2\delta}d\zeta\, \zeta \log\frac{1-\zeta-\rho+2\delta}{1+\zeta-\rho-2\delta}\, ,
\end{equation}
being $8\delta^2$ the measure of the blue diamond domain in the left panel of Fig.~\ref{domains}. While for $\epsilon <0.5+\delta/(1-\rho)$ one can find the following estimation
\begin{equation}
\label{eq:muint2}
\mu_{int,2}=\frac{1}{N_4}\sum_{i\in Z_4}\frac{|\zeta_i|}{1-\gamma_i}\sim\frac{2}{\mathcal{P}}\int_0^{(\zeta_P+\zeta_Q)/2}d\zeta\, \zeta\int_{\zeta+\rho-2\delta}^{-\zeta+\rho+2\delta}d\gamma\frac{1}{1-\gamma}=\frac{2}{\mathcal{P}}\int_0^{(\zeta_P+\zeta_Q)/2}d\zeta\, \zeta \log\frac{1-\zeta-\rho+2\delta}{1+\zeta-\rho-2\delta}\, ,
\end{equation}
where $\mathcal{P}$ is the measure of the polygon $TPQRQ'P'$ (in blue in the right hand side of Fig.~\ref{domains}). Let us observe that in the last integral we make the approximation that points $P$ and $Q$ can be replaced by an average point whose $\zeta$ coordinate is the average of the ones for $P$ and $Q$; this is a minor assumption that helps to compute the integral and will not influence the final result as show below. Now both integrals can be exactly computed.

To get the final estimate for $\langle \mu\rangle$ we need to compute $N_4/N$ and $N_{13}/N$ in the case $\epsilon <0.5+\delta/(1-\rho)$, the other case being trivial and already considered. Assuming the number of points sufficiently large we can affirm that such fractions are well approximated by the ratio of the corresponding polygons, that is
\begin{equation*}
\frac{N_4}{N}\sim \frac{\mathcal{P}}{8\delta^2}\quad\text{and}\quad \frac{N_{13}}{N}\sim 1-\frac{\mathcal{P}}{8\delta^2}\, .
\end{equation*}

In conclusion we can approximate the average magnetization for all $\epsilon>0.5$ with the formula:
\begin{equation}
\label{eq:muapprox}
\langle \mu \rangle^{approx} = \frac{\mathcal{P}}{8\delta^2} \mu_{int,k}+ (1-\frac{\mathcal{P}}{8\delta^2})\, ,
\end{equation}
where $\mu_{int,1}$ is given by Eq.~\eqref{eq:muint1} valid for $\epsilon >0.5+\delta/(1-\rho)$, and $\mu_{int,2}$ by Eq.~\eqref{eq:muint2} for the case $\epsilon <0.5+\delta/(1-\rho)$.

\begin{figure}[h]
\centering
\includegraphics[width=0.8\textwidth]{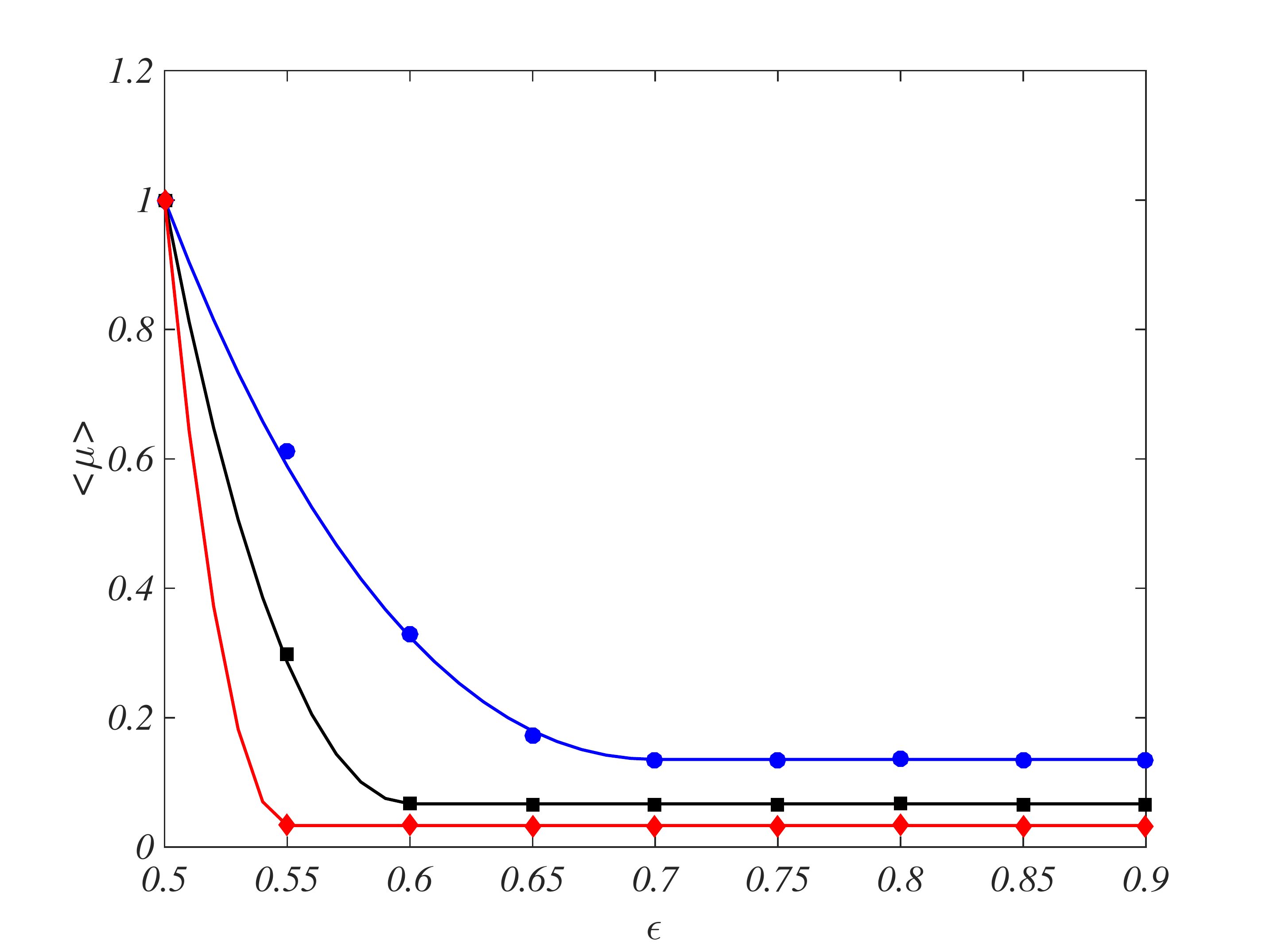}
\caption{\label{comparemu}Averaged magnetization $\langle \mu\rangle$ and the approximation $\langle \mu \rangle^{approx}$ as a function of $\epsilon >0.5$. Symbols correspond to numerical integrations of the system~\eqref{eq:zetagammat} and the use of the definition for the magnetization, lines to the calculation of the approximation Eq.~\eqref{eq:muapprox}. Blue circles correspond to $\delta=0.02$, black squares to $\delta=0.01$ and red diamonds to $\delta=0.005$. Each point correspond to the average over $50$ replicas. The emptiness has been fixed to $\rho=0.9$.}
\end{figure}

In Fig.~\ref{comparemu} we compare the average magnetization obtained using its very first definition Eq.~\eqref{eq:magnezg} and the numerical integration of the system~\eqref{eq:zetagammat}, with the approximation given by Eq.~\eqref{eq:muapprox} for several values of $\delta$ and $\rho=0.9$. One can observe the very good agreement, providing thus an a posteriori validation of the assumptions made so far. Comparing with Fig.~4 of the main text, this provide also a very good approximation to the average magnetization computed using the metapopulation model.

\end{document}